\documentclass[runningheads]{llncs}
\usepackage{graphicx}
\usepackage{subcaption}
\usepackage{amsmath}
\usepackage{amsfonts}
\usepackage{amssymb}
\usepackage{svg}
\begin{document}
\title{Epistemic and Aleatoric Uncertainty Quantification and Surrogate Modelling in High-Performance Multiscale Plasma Physics Simulations\thanks{Supported by organisation ``Munich School of Data Science''}}
\titlerunning{Epistemic and Aleatoric UQ in HPC MS Plasma Physics Simulations}
%
\author{Yehor Yudin\inst{1,2}\orcidID{0000-0001-8867-32073} \and
David Coster\inst{1}\orcidID{0000-0002-2470-9706} \and
Udo von Toussaint\inst{1}\orcidID{0000-0002-8867-1014} \and
Frank Jenko\inst{1,2}\orcidID{0000-0001-6686-1469}}
\authorrunning{Y. Yudin et al.}
%
\institute{Max-Planck-Institute for Plasma Physics, Garching 85748, Germany \and
CS Department, Technical University of Munich, Garching 85748, Germany
\email{\{yehor.yudin, david.coster, udo.v.toussaint, frank.jenko\}@ipp.mpg.de}}
\maketitle              
\begin{abstract}

This work suggests several methods of uncertainty treatment in multiscale modelling and describes their application to a system of coupled turbulent transport simulations of a tokamak plasma.
We propose a method to quantify the usually aleatoric uncertainty of a system in a quasistationary state, estimating the mean values and their errors for quantities of interest, which is average heat fluxes in the case of turbulence simulations.
The method defines the stationarity of the system and suggests a way to balance the computational cost of simulation and the accuracy of estimation.
This allows, contrary to many approaches, to incorporate of aleatoric uncertainties in the analysis of the model and to have a quantifiable decision for simulation runtime.
Furthermore, the paper describes methods for quantifying the epistemic uncertainty of a model and the results of such a procedure for turbulence simulations, identifying the model's sensitivity to particular input parameters and sensitivity to uncertainties in total.
Finally, we introduce a surrogate model approach based on Gaussian Process Regression and present a preliminary result of training and analysing the performance of such a model based on turbulence simulation data.
Such an approach shows a potential to significantly decrease the computational cost of the uncertainty propagation for the given model, making it feasible on current HPC systems.

\end{abstract}
\section{Introduction}


Thermonuclear fusion is a prospective source of clean and abundant renewable energy with a potential for worldwide deployment, independent of geography and politics.
To achieve fusion conditions, one needs to bring plasma temperature and density to sufficiently high values and isolate the plasma from the external matter for a sufficiently long time.
One of the most studied methods to confine a plasma is to do it in a toroidal magnetic trap.
The most developed type of device for confinement is a tokamak, an axisymmetric toroidal chamber with magnetic coils.
One of the critical phenomena interesting for designing and analysing such confinement devices is the distribution of heat and density of the plasma over the minor radius of the torus.
In the latest decades, it was discovered that transport properties of plasmas are dominated by microscopic dynamics of turbulent nature\cite{falchetto_european_2014}.

One of the challenges of such a study is the multiscale nature of the occurring phenomena, with properties of the processes happening across the size of an entire device, of the order of meters, during the confinement relevant times, of the order of seconds, are influenced and driven by microscopic processes, of the order of millimetres and microseconds.

One of the techniques in computational fields applied to tackle such challenges is coupling single-scale models, each solving a subset of equations describing processes happening on a particular scale and being implemented as a separate computer code.
Such codes can, in principle, be run as a standalone program, accepting inputs and producing outputs, adhering to defined data structures.

In this work, we study the simulation of temperature profiles in a tokamak plasma, where plasma properties on the smallest scales define heat transport, and the equations for transport, equilibrium and microscopic turbulence are solved via separate codes, similar to the approach described by Falchetto et al. \cite{falchetto_european_2014}.


The following practical step interesting for scientific purposes is estimating uncertainties in the quantities of interest (QoIs) studied on the largest scale, considering that those uncertainties come from different sources.

To handle uncertainties, a quantity of interest can be treated as a random function of its parameters, also understood as random variables. 
In order to estimate uncertainties of function values using only a finite number of evaluations, assumptions can be made on the function, and an intermediate model or metamodel can be created based on a given sample of evaluations. 
This surrogate model describes the function's dependency on a selected subset of parameters and serves as a replacement for evaluations of the original function for practical purposes.

In this work, we apply surrogate modelling to the turbulent transport plasma simulation, replacing the solution of the microscopic turbulence equation with a fast model able to infer mean values and uncertainties of turbulent heat flux.

\section{Methodology}

\subsection{Epistemic Uncertainty}

One of the most common concerns in science is the uncertainty of inference or predictions due to a lack of knowledge of actual values of specific parameters or lack of knowledge incorporated into the model, in other words, epistemic uncertainty.

In theoretical fields, particularly in computational areas, mathematical models often involve unknown parameters that can only be estimated probabilistically based on prior knowledge, observations or experiments. 
The uncertainty propagation problem means finding an approximation of the probability density function of quantities of interest in the model, given a set of uncertain parameters that are understood as random variables with known probability density functions.

As it is often difficult to retrieve a particular parametric form of QoI PDFs, the useful statistics describing the properties of output random variables are the statistical moments of the density.

\subsubsection{Polynomial Chaos Expansion }

One of the most commonly used methods for uncertainty quantification is Polynomial Chaos Expansion (PCE)\cite{sullivan_uq_2015}.
The core idea of the method is to expand the function of output $Y$ of inputs $X$ using the finite basis of orthonormal polynomials $P_{i}(x)$ up to the order of $p$, each depending on a subset of input components and being orthogonal to their PDFs:
\begin{equation}
Y \approx \hat{Y}(X) = \sum_{i=0}^{N} c_{i}P_{i}(X)
\end{equation}

For the function $Y=f_{X}(x)$, the expansion coefficient $\omega_{k}$ and the ordinate values $x_{k}$ could be determined via a quadrature scheme based on a Spectral Projection method exploiting the orthogonality of the basis polynomials and using normalisation factors $H_{i}$ for polynomials $P_{i}$:

\begin{equation}
 c_{i} = \frac{1}{H_{i}} \int_{\Omega} Y(x)P_{i}(x)f_{X}(x)dx \approx \frac{1}{H_{i}} \sum_{k=1}^{N} Y(x_{k})P_{i}(x_{k})\omega_{k}
\end{equation}

The benefit of such an orthonormal approach is that the expansion coefficients are sufficient to calculate the estimates of the moments of the output PDF.

\begin{equation} 
\hat{\mathbb{E}}[Y] = c_{1} ,\quad \hat{\mathbb{V}}[Y] = \sum_{i=2}^{N} \mathbb{E}[P_{i}^{2}(X)] c_{i}
\end{equation}

However, this method suffers from the curse of dimensionality as the number of required samples grows exponentially, like a binomial coefficient, with the number of dimensions $d: N=(d+p)!/p! \cdot d!$ 

\subsubsection{Sensitivity Analysis}

A particular type of statistics over a sample of code's QoI responses aims to define the contribution of a single parameter, or a subset of parameters, to the uncertainty of the output. 
In this work, we study Sobol indices\cite{sobol_sensitivity_1990}, which is a global variance-based sensitivity metric.
The quantity we estimate is the variance of the mathematical expectations of the function output, conditioned on a subset of input components $\{i\}$ and normalised by the total variance of the output.
Here we use the first-order Sobol indices, identifying the influence of a single parameter $i$; higher-order Sobol indices, determining the influence of a subset of interacting parameters $\{i\}$; and total Sobol indices, identifying the total effect of each parameter, using a set of all components $-i$, excluding a single $i$ component:
\begin{equation} 
S_{\{i\}} = \frac{\mathbb{V}[\mathbb{E}[Y|X_{\{i\}}]]}{\mathbb{V}[Y]} \, , \quad S_{i}^{tot} = 1- \frac{\mathbb{V}[\mathbb{E}[Y|X_{-i}]]}{\mathbb{V}[Y]} 
\end{equation}

Using such an indicator, one could judge which parameters are most important for the behaviour of the computational model and decide whether a particular parameter's value can be fixed in subsequent studies or whether more values should be tried out.

\subsection{Aleatoric Uncertainty}

One of the types of uncertainties, different from the epistemic, is the aleatoric uncertainty coming from the inherent properties of the model and its solver. 
In the first place, it is related to the dynamic stochastic behaviour of the quantities of interests, for which the model is solved.
Several methods for analysing aleatoric uncertainties commonly involve two principles: separating epistemic uncertainties and performing statistical analyses of output quantities. 
One approach uses replica sampling, in which a certain number of model replicas are solved for each set of input parameters, followed by analysing the distributions of quantities of interest values\cite{vassaux_ensembles_2021}. 
Other methods are suitable for analysing the quasistationary nonlinear behaviour of the model in the vicinity of a certain attractor over a considerable time of solution\cite{brajard_combining_2021}.

This work utilises a practical measure for an aleatoric uncertainty based on analysis of the standard error of the mean values of quantities of interest $E=\mathrm{SEM}[y]=\sigma[y]/\sqrt{n_{\mathrm{eff}}}$ for a sample of observations $\mathbf{Y}^{\mathrm{eff}}$, each taken in an autocorrelation time window of length $t_{\mathrm{A}}$ apart from the previous one.
Such an approach allows judging the requirements of a single solution for a given set of parameters and controlling the level of aleatoric uncertainty.
However, such a measure only provides information on the second moments of the quantity of interest distribution. 
Furthermore, it is not always well incorporable into a single framework of analysis of both epistemic and aleatoric uncertainties.

\subsection{Surrogate Modelling}

This work primarily focuses on applying Gaussian Process Regression (GPR), a non-parametric probabilistic machine learning model fully defined by the covariance matrix of observed data samples and capable of providing a posterior PDF of output given any input values from the considered support\cite{rasmussen_gpr_2006}.

When the Matern-$3/2$ kernel $k$ describes covariance between any two parametric points $x_i, x_j$, determining a PDF $p(f(x_{\ast}))$ of a posterior for new data $x_{\ast}$ using a standard GPR based on a sample of observed inputs $X$ and outputs $y$ with a Normal likelihood looks like following:
\begin{align}
        & p(f(x_{\ast})|\mathbf{X},\mathbf{y},x_{\ast}) \sim \mathcal{N}(\mu(x_{\ast}), \sigma^{2}(x_{\ast})) \\
        & \sigma^{2}(x_{\ast}) = K(x_{\ast},x_{\ast})- K(x_{\ast},\mathbf{X})^{\top} \left(K(\mathbf{X},\mathbf{X})+\sigma_{e}^{2}\mathbf(I)\right)^{-1}K(x_{\ast},\mathbf{X}) \\
        & K=\left( k(x_{i}, x_{j}) \right)_{i,j=1..N} \\
        & k(x_{i},x_{j}) = \sigma^{2}\left(1+\sqrt{3}\frac{|x_{i}-x_{j}|}{l}\right)\exp\left(-\sqrt{3}\frac{|x_{i}-x_{j}|}{l}\right)
\end{align}

The method allows for capturing uncertainties in the underlying data that come from the lack of information on the parametric dependencies and internal noise.
Furthermore, this method has a high prediction uncertainty in the region of the parametric space for which there was not enough training data, as the posterior would have a higher variance for points decorrelated with the observed sample.
Such a high uncertainty indicates the untrustworthiness of the regression results and the model's limitations.


One of the ideas behind the surrogate modelling approach is to create flexible and adaptive surrogates based on new data. 
Online Learning allows adding new data into the training sample to improve the surrogate gradually used as a proxy solver\cite{leiter_accelerated_2018}. 
Active Learning uses the regression model's prediction uncertainty and information about sought dependency to determine new points in the parameter space for the evaluation of the function of interest\cite{preuss_global_2018}.

In this work, we apply Gaussian Process Regression trained on a data set describing performed turbulence simulations.
We trained such a surrogate model for the dependency of heat fluxes on the temperature values and its gradient for a particular configurational location in a tokamak plasma.

\section{Numerical Results}

\subsection{Computational Model}

\begin{figure}
    \centering
    \includegraphics[width=0.65\textwidth]{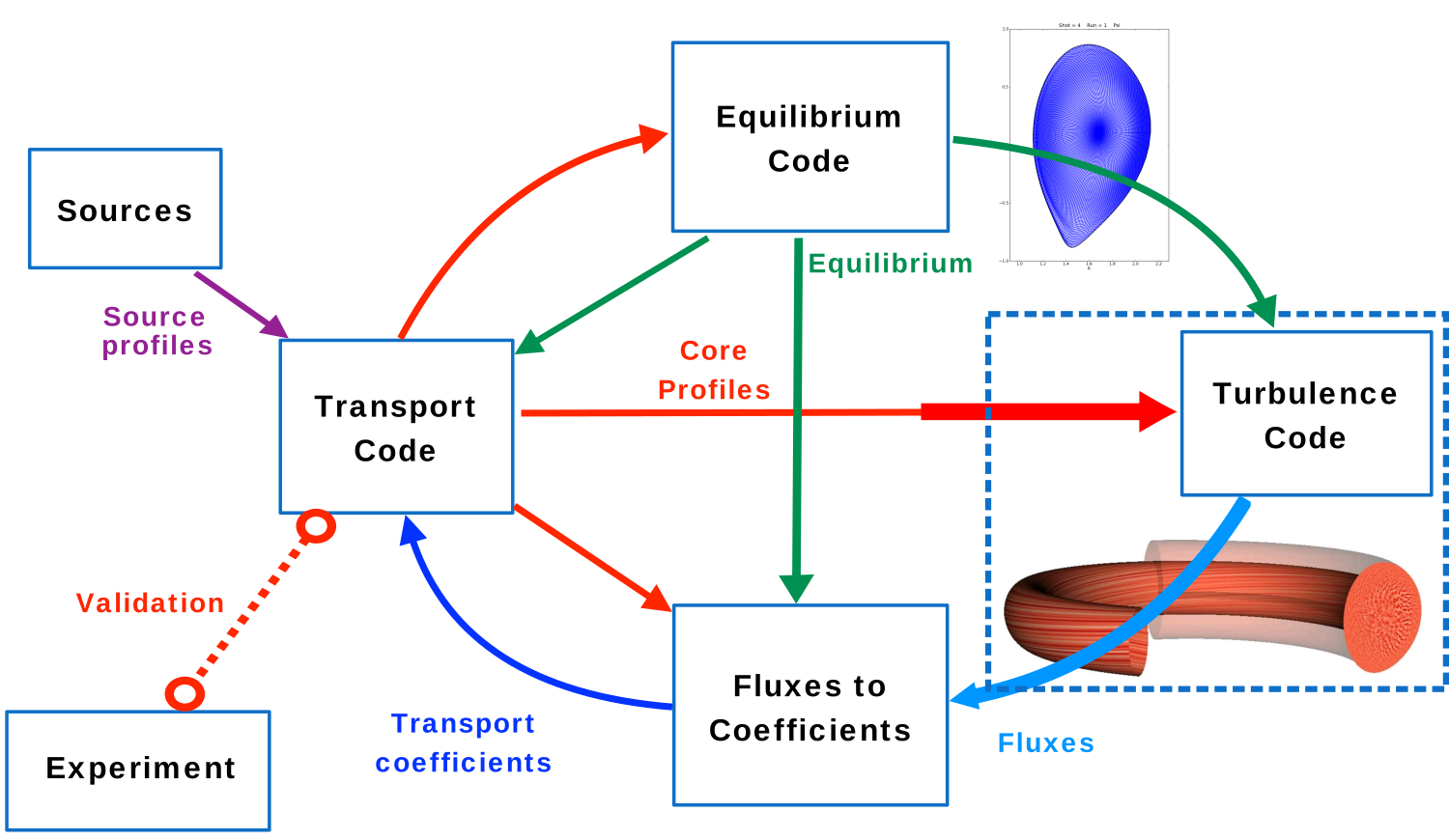}
    \caption{Main components of the Multiscale Fusion Workflow. 
    The turbulence model is the one that requires the most computational resources to solve, and that exposes the highest sensitivity to its inputs.} \label{fig_mfw}
\end{figure}

The object of the study is the effects of turbulent processes happening at microscopic time and space scales on the temperature profiles of a nuclear fusion device across the confinement scales.
For that, we use the Multiscale Fusion Workflow (MFW)\cite{luk_compat_2019} consisting of three main models, each implemented as a separate computer code, serving as an independent workflow component capable of exchanging its solution with others in a black-box fashion.
The models are the equilibrium code that describes plasma geometry, the transport code that evolves the plasma profiles and the turbulence code used to compute effective transport coefficients (Fig \ref{fig_mfw}).
The code that solves equilibrium Grad-Shafranov equations for plasma density and the magnetic field is CHEASE\cite{luetjens_chease_1996}, a high-resolution fixed-boundary solver. 
Evolution of temperature and density is performed with ETS\cite{coster_european_2010}, a 1D code for energy and particle transport.

In this work, we analyse the turbulence component of the workflow, for which we use the 3D nonlinear gyrofluid turbulence code GEM\cite{scott_free-energy_2005} in its local flux tube approximation.
Such turbulence models are very sensitive to input parameters and computationally expensive to solve. 
Thus, it is a prime point for performing a study of uncertainties in the joint multiscale model and the component for which a fast and easy-to-evaluate surrogate model would be of interest to create.
As discussed in \cite{coster_building_2021}, performing UQ on the MFW with GEM is currently computationally unreasonable since 100~000 core hours are required for a single run, whereas a surrogate model should be at least 10~000 times cheaper (1 core instead of 1024, and a factor of 10 per cycle around the MFW loop).

Codes store their respective solution in standardised data structures of Consistent Physical Objects (CPO)\cite{imbeaux_generic_2010}, and the turbulence quantities of interests are stored in a structure describing core transport properties.
In principle, the workflow implemented to analyse the flux time traces and prepare a dataset for surrogate training and utilisation can support any code capable of interfacing via CPOs.

The resulting data structures are passed from a code to a code instance in a point-to-point fashion using the MUSCLE2 coupling library\cite{borgdorff_distributed_2014}.
Currently, a version supporting various turbulence models is being implemented using the MUSCLE3 library\cite{Veen2020}.

\subsection{Simulations and resulting time traces}

A number of simulations of turbulence code for different input parameter values were performed to study the behaviour of quantities of interest in turbulence.
Furthermore, this data was used to analyse what information can be included in a surrogate model and as an initial data set for the model training.

Transport equations describe temperature and density behaviour using a 1D approach, where quantities are functions of radial coordinate and time but must be solved globally for the entire domain. 
Turbulence occurs on a smaller scale, and behaviour around a single magnetic flux tube or a small range of the radial domain typically defines it very well.
In order to interpolate the turbulence solution for the entire device, the tokamak turbulence model for the ASDEX Upgrade\cite{falchetto_european_2014} must be solved for at least eight different radial positions\cite{coster_building_2021}.

The turbulent model solution's quantity of interest is the heat flux averaged across a magnetic field surface at a specific radial position in the toroidal device. 
The computed result is nonlinear on microsecond timescales, with several features and stages typical for all relevant parameter space. 
Initially non-turbulent, the effective QoI value grows exponentially as turbulence develops until nonlinear effects dominate, causing turbulence to saturate and QoI to behave quasistationary. 
Although mean values remain around a particular mean value, the QoI is also locally chaotic and fluctuates substantially over time.

Defining whether the system is saturated, accurately representing the mean output values for a particular case, and decoupling fluctuating behaviour are crucial to creating an effective surrogate model parametrising the transport properties.

Macroscopic kinetic profiles behave similarly to micro-level turbulent fluxes over confinement-relevant times. 
Starting from initial conditions, they converge towards a stationary level self-consistent with the coupled model.
The profiles dynamically and nonlinearly interact with local turbulent behaviour, resulting in a quasistationary solution. 
In order to capture the dependency of turbulent fluxes in relevant parameter space, a surrogate model must be based on a large sample of code solutions covering the relevant regions.

Creating a sample for surrogate training involves running the same code for multiple input parameter values and managing the simulation and training data, which pose a significant challenge.
Using the resulting surrogate as a microscopic code replacement requires ensuring it captures QoI dependency well for all practically possible parameter values. 
This requires a training dataset covering enough parameter space information; the surrogate should indicate its applicability limits and epistemic uncertainty due to a lack of training data.

Given the practical challenges, this work suggests another novel computational workflow besides one of the multiscale simulations to solve the turbulent transport problem.
The new workflow analyses turbulent fluxes time traces on the microscopic temporal scales and defines the stationarity of the QoI as well as its mean values and estimation error.

\subsection{Aleatoric uncertainty}

One of the work's challenges was to separate fundamental irreducible uncertainty from epistemic parametric uncertainty in a fluctuating system. 
It was important to establish the stationary mean values for analysing key parametric dependencies and to use fluctuations only for statistical error estimation. 
However, solving the model for a long enough time to gather sufficient data is computationally demanding and gathering a large sample would be prohibitively expensive. 
Estimating error levels during the course of a simulation run can help decide when to stop the simulation and significantly save computational costs.


The procedure for analysing the QoI time traces $y(t)$, representing a model solution for a single parametric point, constitutes of the following steps:
\begin{enumerate}
    \item For the time-traces $y(t)$ of length $t_{n}$ representing a model solution in a scalar quantity of interest at time steps $\{t\}$, we select a part in the saturated phase:
    \begin{enumerate}
        \item Define the $y(t)$ ramp-up phase duration $t_{\mathrm{r.u.}}$: here, in practice, for the long term, we chose an initial $15\%$ of readings 
        \item Discard the readings from the ramp-up phase
    \end{enumerate}
    \item Downsample the readings:
    \begin{enumerate}
        \item Calculate the Auto-Correlation Time:
        $t_{\mathrm{A}}=\mathrm{ACT}[y(t)]=\mathrm{min}\, t^{\ast} : \\ \frac{1}{(t_{n}-t^{\ast})} \sum_{t=t^{\ast}}^{t_{n}} \left(y(t)-\bar{y}\right)\cdot \left(y(t-t^{\ast}\right)-\bar{y}) < \frac{1}{t_{n}} \sum_{t=t_1}^{t_{n}} y^{2}(t) \cdot e^{-1} $
        \item Split time series in saturated phase into $n_{\mathrm{eff}}^{}=\lfloor \frac{t_{n}}{t_{\mathrm{A}}} \rfloor $ windows
        \item For every autocorrelation time window of length $t_{\mathrm{A}}$ choose a downsampled reading as a mean value $y_{i}^{\mathrm{eff}}=\frac{1}{t_{\mathrm{A}}}\sum_{j=i \cdot t_{\mathrm{A}}}^{(i+1) \cdot t_{\mathrm{A}}} y_{j}$ for an effective time step $t_{i}^{\mathrm{eff}}=\frac{1}{t_{\mathrm{A}}}\sum_{j=i \cdot t_{\mathrm{A}}}^{(i+1) \cdot t_{\mathrm{A}}} j$
        \item Collect downsampled readings into a new set $\mathbf{Y}^{\mathrm{eff}}=\{y_{i}^{\mathrm{eff}}\}$
    \end{enumerate}
    \item Test the stationarity of the resulting time series:
    \begin{enumerate}
        \item Here: compute an ordinary least-squares linear multivariate regression model of downsampled QoI readings over the time
        \item Apply Normal Equations to find coefficients: $\hat{\mathbf{\theta}} = \left(\mathbf{X}^{\top}\mathbf{X}\right)^{-1}\mathbf{X}^{\top} \mathbf{Y}$ where $\mathbf{X}$ consists of effective time steps $t_{i}^{\mathrm{eff}}$
        \item Test if the linear regression coefficients are below a chosen relative tolerance: $\hat{\mathbf{\theta}} < \mathbf{\epsilon}_{\mathrm{tol}}$
        \begin{enumerate}
            \item If the coefficients are too large, continue running the simulation for another $t_{\mathrm{run}}$ time steps
            \item If the coefficients are small enough, stop the simulation and proceed to the statistics calculation
        \end{enumerate}
    \end{enumerate}
    \item Compute the essential statistics estimates:
    \begin{enumerate}
        \item Mean:  $\mu[y] = \frac{1}{n_{\mathrm{eff}}^{\mathrm{}}}\, \sum_{y\in \mathbf{Y}^{\mathrm{eff}}} y $
        \item Standard deviation: $\sigma [y] = \left(\frac{1}{n_{\mathrm{eff}}^{\mathrm{}}-1} \, \sum_{y \in \mathbf{Y}^{\mathrm{eff}}} (y-\mu[y])^{2}\right)^{-\frac{1}{2}}$
        \item Standard Error of the Mean: $\mathrm{SEM}[y] = \sigma [y] / \sqrt{n_{\mathrm{eff}}^{\mathrm{}}} $
    \end{enumerate}
\end{enumerate}

We applied this procedure for the time traces produced by turbulence code iteratively, in particular to ion heat flux values over time, sampling after a number of solution time steps, and analysed the behaviour of the mean estimate and its error.

\begin{figure}
    \centering
    \includegraphics[width=0.75\textwidth]{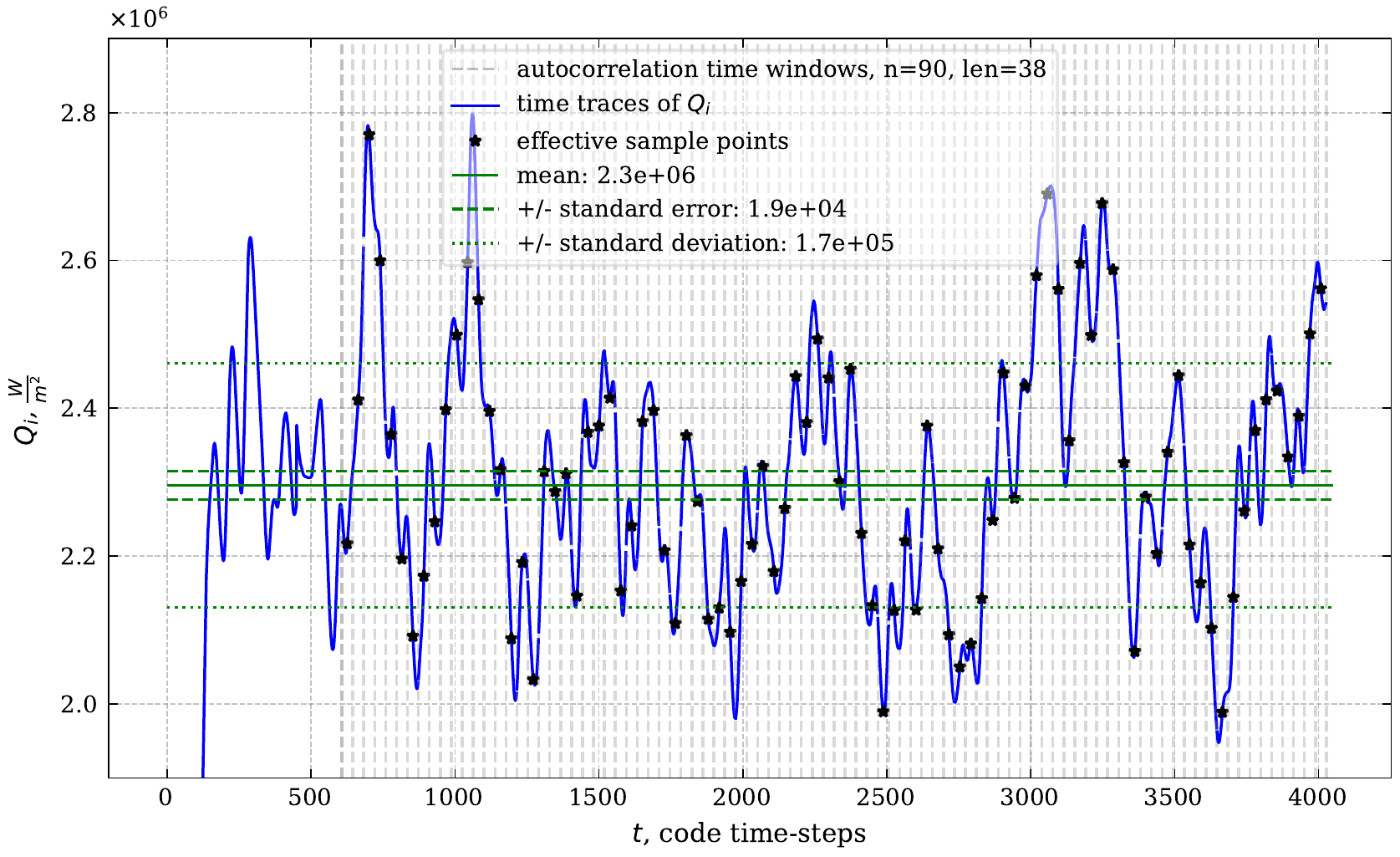}
    \caption{The standard error of the mean is calculated accounting to discarding 15\% of reading that do not belong to stationary phase, calculating the autocorrelation time of the remaining time traces, taken a sub-sample constituent of a single mean reading per autocorrelation window and renormalising mean error using the effective size of the sample.} \label{fig_traces_1}
\end{figure}

Figure \ref{fig_traces_1} describes mean values calculated with the sequential procedure based on the data of time traces of heat fluxes in a quasistationary phase.

\begin{figure}
\centering
    \includegraphics[width=0.7\textwidth]{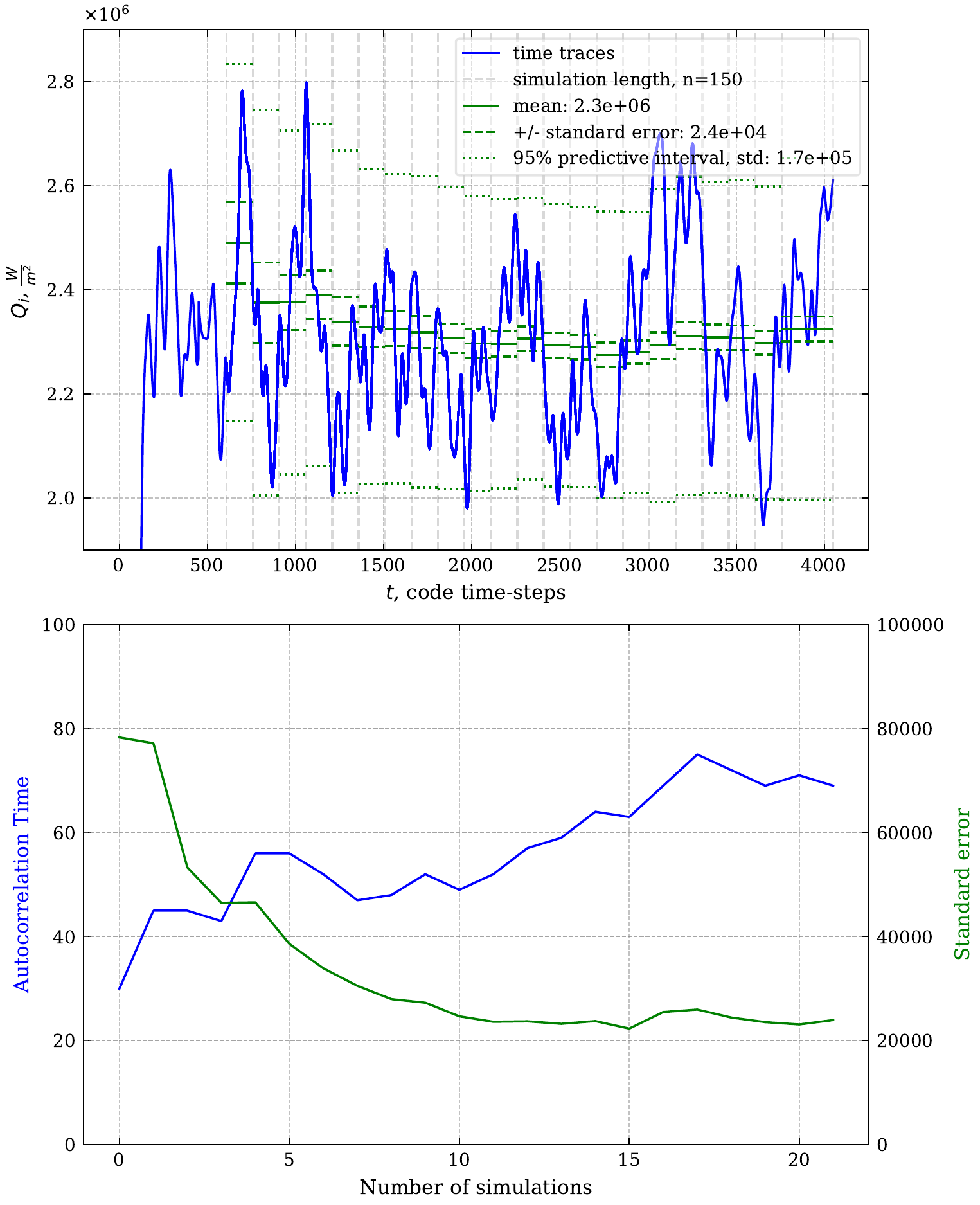}
\caption{Sequential analysis of time traces with an interval of 150 time-steps per window, defined by 6 hours of computation on a single MPCDF's COBRA node.
After each step of time traces computation, we recalculate the autocorrelation time, repopulate an effective data sample, and estimate new values for the mean, standard deviation and standard error of the mean.
With the growth of the sample, the mean estimation converges to the true value, and the standard error of the mean decreases to around $1\%$ of the respective mean value after $2300$ time steps out of $4000$ in total.
Here, one can decide to stop the computation using a tolerance threshold on the relative change of the mean estimate of $10^{-3}$ or by the error of the mean dropping below $1.1\%$ Coefficient of Variation.
The plots of the standard error of the mean and autocorrelation time dependency on number simulation steps indicate a convergence of the time traces analysis with simulation time.
} \label{fig_traces_2}
\end{figure}

After the relative change of the mean and the relative change of the standard error of the mean converge below a certain threshold, as shown in Figure \ref{fig_traces_2}, we considered the duration of a simulation sufficient enough to have a reasonable estimate of the average flux and its error and added this data to a dataset describing the computational turbulence model behaviour.
Subsequently, this dataset is used to draw conclusions on the model's parametric uncertainties and train a surrogate model.

\subsection{Epistemic Uncertainty}

To explore the epistemic uncertainties of the properties of the turbulence model, we performed an uncertainty propagation, taking a particular input of macroscopic parameter with a value taken from experimental measurement and, assuming that on top of this value, there is uncertainty, described as a normally distributed random variable with a Coefficient of Variation, a ratio of the variance of a quantity to its mean value, $\mathrm{CV}(x)=\sqrt{\mathbb{V}[x]}/\mathbb{E}[x]$ equal to $0.1$. 
The statistics for average ion and electron heat flux $\mathbf{y} = (Q_{i}(\rho), Q_{e}(\rho))$ were estimated using a PCE method with Hermite polynomials of order $p=2$ applied for four input parameters of ion and electron temperatures profile values and gradients: $\mathbf{x}=T_{i}(\rho), T_{e}(\rho), \nabla T_{i}(\rho), \nabla T_{e}(\rho)$.

The uncertainty propagation was performed using EasyVVUQ\cite{easyvvuq2020} library, which manages the definition of input uncertainties, sampling scheme, generation of input data structures and collection of the output ones, as well as calculating statistics.
The batches of the turbulence code instances were run on the MPCDF COBRA supercomputer using QCG-PilotJob\cite{piontek_development_2016} middle-ware library, which manages HPC resources of a single computational allocation and distributes it to run multiple instances of the code with a particular parallelisation set-up.

One of the issues of uncertainty propagation through the model that exposes chaotic behaviour on some scales is to untangle the epistemic and aleatoric uncertainties.
Before analysing the parametric uncertainty propagation results, the time traces produced by the simulations are processed as described in the previous subsection.

\begin{figure}
    \centering
    \begin{minipage}{0.46\textwidth}
        \centering
        \begin{subfigure}{\linewidth}
            \centering
            \includegraphics[width=1.\linewidth]{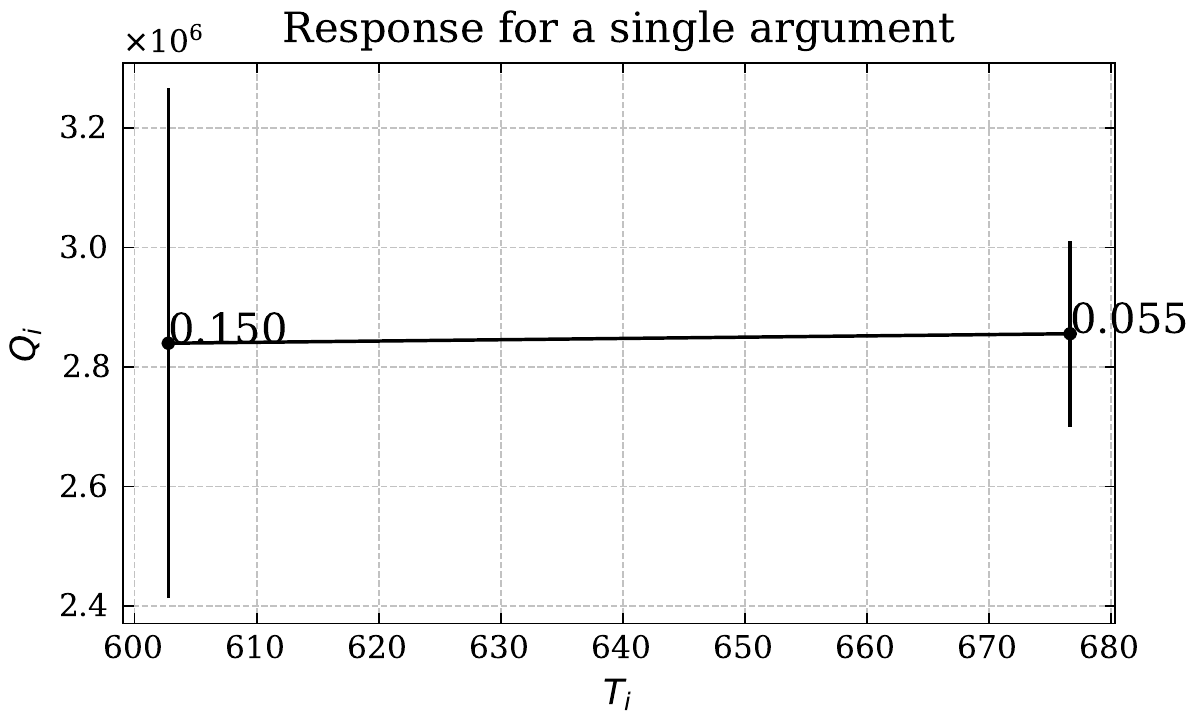}
            \label{fig_meandep_01cov}
        \end{subfigure}
    \end{minipage}
    \begin{minipage}{0.46\textwidth}
        \centering
        \begin{subfigure}{\linewidth}
            \centering
            \includegraphics[width=1.\linewidth]{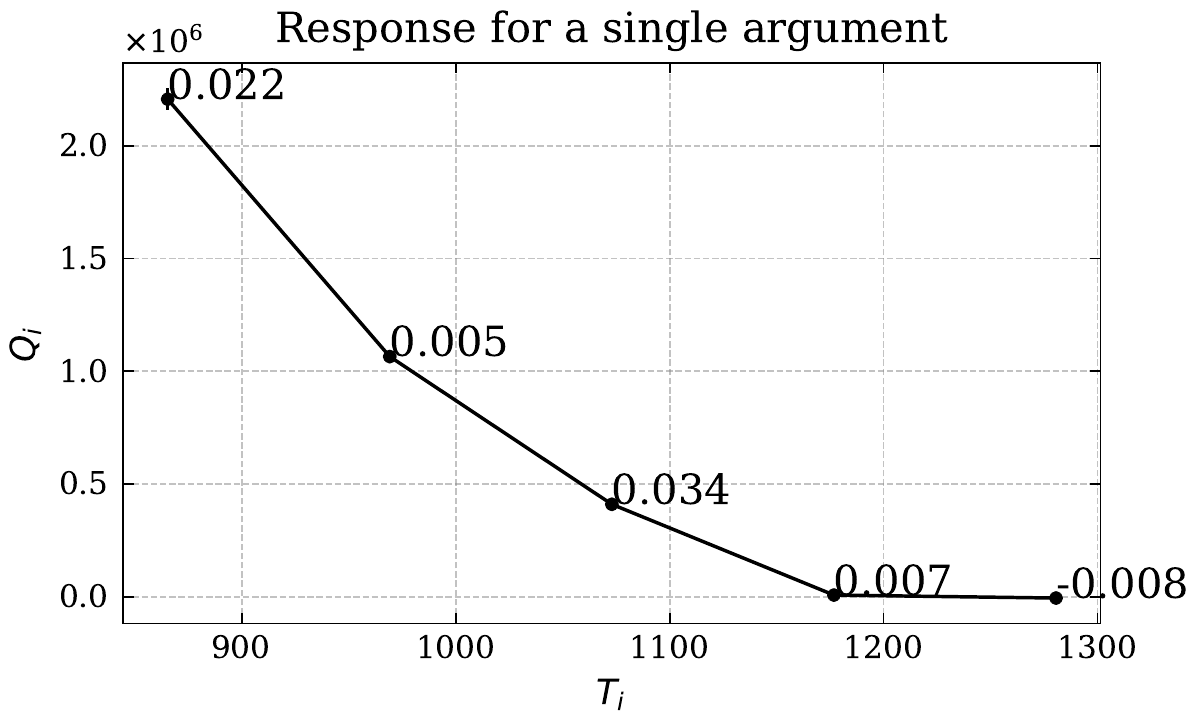}
            \label{fig_meandep_025cov}
        \end{subfigure}
    \end{minipage}
\caption{Dependency of the mean ion heat flux on the ion temperature. For the small variation of input parameters, the mean estimation could not allow for drawing good conclusions on the gradient of this dependency, and the means are within uncertainty intervals of neighbouring points. For the larger variation of input parameters, the relative errors of the mean, also stated next to the points, are small enough to reconstruct the dependency with sufficient certainty.}\label{fig_meandep}
\end{figure}

\begin{figure}
    \centering
    \begin{minipage}{0.42\textwidth}
        \centering
        \begin{subfigure}{\linewidth}
            \centering
            \includegraphics[width=0.9\linewidth]{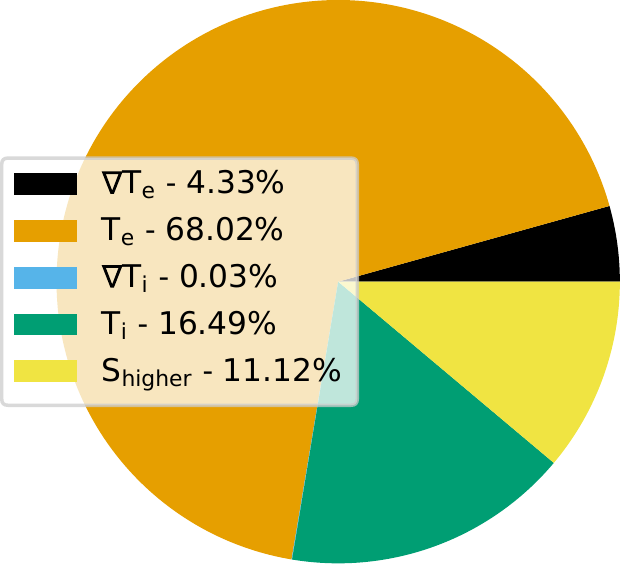}
            \label{fig_sobol_te}
        \end{subfigure}
    \end{minipage}
    \begin{minipage}{0.42\textwidth}
        \centering
        \begin{subfigure}{\linewidth}
            \centering
            \includegraphics[width=0.9\linewidth]{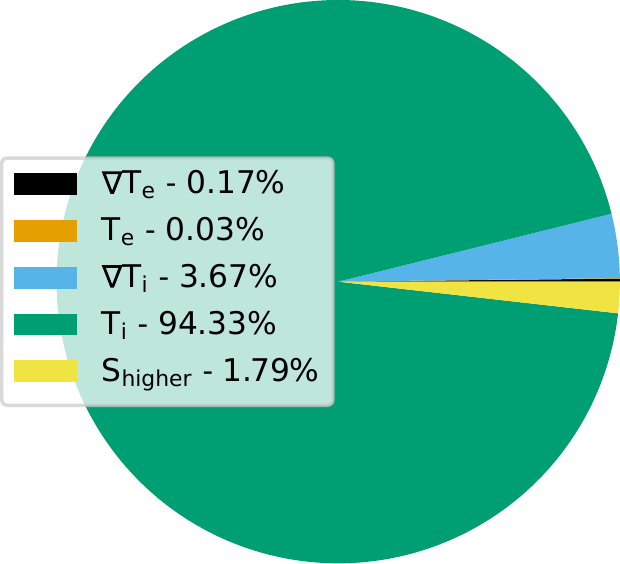}
            \label{fig_sobol_ti}
        \end{subfigure}
    \end{minipage}
\caption{Sobol indices for electron and ion heat fluxes. Most of the variation in electron heat flux is produced by the variation in the electron temperature. Interactions of the input parameters explain some of the non-negligible variation. For the ion heat flux, the output variation is dominated by a single parameter in this regime. However, a fraction is still explained by the nonlinear interaction of parameters.}\label{fig_sobol}
\end{figure}

The result of this study with an error of $\mathrm{CV(\mathbf{x})}=0.1$ was inconclusive, as there was a significant overlap of the error bounds for mean values of quantities of interest for neighbouring points in parametric space (Fig. \ref{fig_meandep}). 
In this case, it was impossible to conclude with certainty whether the ion heat flux trend was rising or falling with respect to the growth of ion temperature and its gradient.
As the next step, we have taken an input parameter with normal distribution with $\mathrm{CV(\mathbf{x})} = 0.25$ and propagated this uncertainty through the model.
Enlarging the input variation allowed us to study the behaviour and uncertainties of the model over the broader range of input parameter values and recover the character of the dependency of ion heat flux on ion temperature and its gradient (Fig. \ref{fig_meandep}). 
Furthermore, longer simulation runs allowed reaching lower levels of the aleatoric uncertainty estimates in the quantities of interest,

The uncertainty analysis results per input component show a dominant parameter for which the model is most sensitive (Fig. \ref{fig_sobol}).
For this particular prior PDF over parametric space, the behaviour of the electron heat flux is most sensitive to variation in electron temperature. Still, other parameters and their interactions also play a significant role.
Other metrics that were chosen to analyse the model's epistemic uncertainty are the Coefficient of Variation in the quantity of interest and the Uncertainty Amplification Factor\cite{edeling_impact_2021}.
In this work, the coefficient of variation of the model's output ion heat flux is about $\mathrm{CV}(Q_{i}) \approx 2.96$, showing a significant range of possible values that QoI can take in the study.

The Uncertainty Amplification Factor is the ratio of the Coefficient of Variation of a QoI to the coefficient of variation of the input parameter $\mathrm{UAF}(y|x) = \mathrm{CV}(y)/\mathrm{CV}(x)$.
In the case of the studied model and chosen input parameters prior distribution, the Uncertainty Amplification Factor of ion heat flux is $\mathrm{UAF}(Q_{i}|\mathbf{x}) \approx 11.9$ showing a significant sensitivity of the model to the parametric uncertainty.

\subsection{Surrogate model}

Having performed the original epistemic uncertainty propagation, we have collected a substantial sample of the physical model's input-output pairs.
This sample was used to fit a surrogate model able to capture the behaviour of the physical model QoI means values and uncertainties over the selected input parametric subspace.
Here, the QoIs and the input parameters are the same as described in the epistemic uncertainty propagation results section.
\begin{figure}
    \centering
    \includegraphics[width=0.75\textwidth]{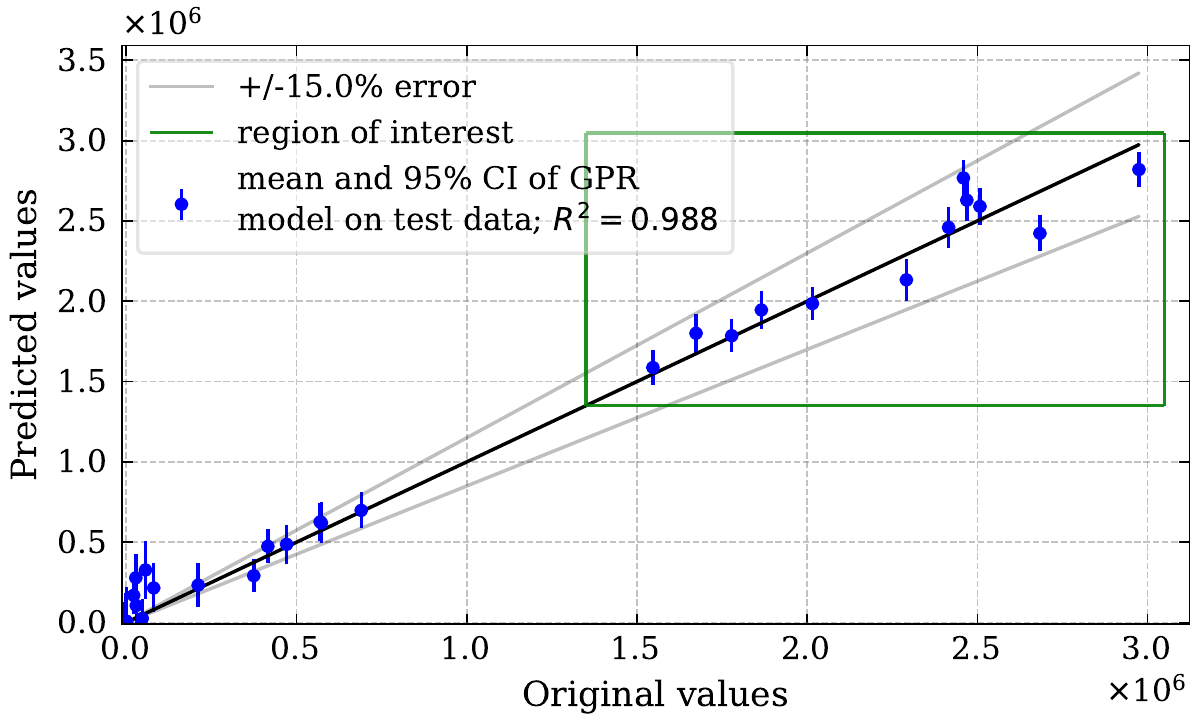}
    \caption{GPR surrogate model test results with the coefficient of determination $R^{2}=0.988$.
    Inferred values of ion heat flux compared to the observed values produced by the physical model. 
    Vertical bars denote the $\pm1.96$ of the standard deviation of the predictive posterior distribution, grey bounds denote 15\% relative error, and dotted lines denote the region of interest.
    For most test samples, the true value is within $95\%$ prediction interval, and the predicted mean is within $15\%$ relative error.} \label{fig_surr_compar}    
\end{figure}

The GPR surrogate provides an intermediate model based on the assumptions on the regularity of the actual dependency, primarily encoded in the covariance kernel and the likelihood of the regression model.
Unlike the PCE surrogate, which depends on the predefined polynomial order and is difficult to adapt given new information about the sought dependency, the uncertainties of the GPR one produce strong indications that the information on the actual dependency is sufficient for the regression and is easily adaptable through sequential design and active learning algorithms.

Here we present the results of the Gaussian Process Regression surrogate trained on $50\%$ of the collected simulation dataset (Fig. \ref{fig_surr_compar}).
We developed functionality for the training and analysis of surrogates based on EasySurrogate\cite{edeling_easysurrogate_2023} library.
After performing a hyper-parameter optimisation, a particular set of categorical parameter values was chosen to judge surrogates by the highest coefficient of determination of QoI values predicted by a surrogate $y_{i}^{\mathrm{surr}}$ compared to the ground truth simulation values $y_{i}^{\mathrm{g.t.}}$ for the model testing: $R^{2} = 1 - {\sum_{i=1}^{n_{\mathrm{test}}}\left(y_{i}^{\mathrm{g.t.}} - y_{i}^{\mathrm{surr}}\right)^{2}} / {\sum_{i=1}^{n_{\mathrm{test}}}\left(y_{i}^{\mathrm{g.t.}} - {\overline{y^{\mathrm{g.t.}}}}\right)^{2}}$


The future plan is to apply a validation test for the applicability of such a surrogate as a replacement for the solution of the turbulence model for given inputs describing core profiles and inferring the average heat fluxes and effective transport coefficients.
The original simulation workflow utilising the turbulence code GEM required around $27.0$ s on $1024$ cores of Cineca MARCONI supercomputer\cite{luk_compat_2019} to predict fluxes for the next iteration of core profiles advancement, whether the EasySurrogate GPR model requires a fraction of a second on a single core for the same procedure.
Furthermore, applying a GPR model would allow the detection of points in parameter space where the surrogate uncertainty is too high and should not be applied as a proxy for a high-fidelity solver.
When it is understood that a GPR surrogate performs sub-par for a region of interest in parametric space, some work was done on designing an Active Learning scheme based on Bayesian Optimisation and Sequential Design of Experiments.

\section{Discussion}

In this paper, we presented a number of methods to treat uncertainties in multiscale simulations. 
This includes ways to measure aleatoric and epistemic uncertainties, to control and reduce the amount of computational resources required to quantification of the uncertainties, as well as to identify situations when uncertainties are hindering analysis of the physical model behaviour.
The suggested method for aleatoric uncertainty processing is applicable when some model dependency could be decomposed into the mean and fluctuation behaviour and allows for a trade-off between the computational effort for the model solution and the accuracy of the mean behaviour estimation.
The surrogate modelling method proposes a way to make epistemic uncertainty propagation cheaper by constructing a data-based machine learning GPR proxy model for an expensive physical model solver that could be used as a replacement in multiscale coupled simulations and for sampling in uncertainty quantification methods.

We applied the described methods for the dependency of average heat fluxes produced by plasma turbulence code GEM based on tokamak core temperature profiles and their gradients.
While studying such fluxes' time traces, we identified cases when the simulations converged early enough to allow reducing the duration of runs and saving computational resources.
The QoI mean levels were estimated with sufficient certainty to draw conclusions about sought dependency and use solution data as a basis for further analysis and surrogate model training.
The subsequently performed uncertainty propagation and sensitivity analysis showed situations where aleatoric uncertainties could not allow concluding on parametric dependencies.
Also, we defined cases with a strong indication that a particular set of variables influences output variation dominantly.
The surrogate modelling approach of fitting a Gaussian Process Regression model to the dataset of flux computing turbulent simulations showed that it is possible to train a metamodel that allows capturing the mean behaviour of the fluxes well within the posterior predictive bounds for a region of interest in parameter space of the model.

The development of a surrogate for the turbulence code opens up the much-needed possibility of doing detailed UQ for the whole MFW workflow with the accuracy of the underlying turbulence code but without the expense of the turbulence code in the workflow.

Future work would include testing the surrogate as a proxy of the expensive turbulence code, capable of inferring fluxes and subsequently defining the effective transport coefficients.
This includes implementing an expanded version of a Multiscale Fusion Workflow, capable of switching between different implementations of the turbulence model and validating the results of the coupled turbulent transport simulations using a turbulence surrogate against a version using the gyrofluid code.
Subsequently, such a modified workflow should include capacity for adaptive surrogate training through Online Learning and Active Learning algorithms, as well as methods to select turbulence model implementation in the course of a single simulation.

\section{Acknowledgements} 
The authors of this paper would like to acknowledge the support of the Poznan Supercomputer and Networking Centre (PSNC) and Max-Planck Computational Data Facility (MPCDF).
Research by Yehor Yudin is funded by the Helmholtz Association under the "Munich School for Data Science - MuDS".
%
%
%
%


\bibliographystyle{splncs04}
\bibliography{ref}


\end{document}